\documentclass[conf, letter, 10pt]{IEEEtran}
\usepackage{amsmath}
\usepackage{amsfonts}
\usepackage{amssymb}
\usepackage{amsthm}
\usepackage{commath}
\usepackage{graphicx}
\usepackage{bm}
\usepackage{bbm}
\usepackage{multirow}
\usepackage{epstopdf}
\usepackage{physics}
\usepackage{mathtools}
\usepackage{cases}
\usepackage{enumitem}
\usepackage{cite}
\usepackage{xcolor}

\usepackage{caption}
\usepackage{subcaption}
\allowdisplaybreaks
\newtheorem{theorem}{Theorem}
\newtheorem{definition}{Definition}
\newtheorem{lemma}{Lemma}
\newtheorem{proposition}{Proposition}

\theoremstyle{definition}
\newtheorem{remark}{Remark}
\theoremstyle{definition}
\newtheorem{example}{Example}
\usepackage[top=0.6in, bottom=0.95in, left=0.62in, right=0.62in]{geometry}

\newcommand\scalemath[2]{\scalebox{#1}{\mbox{\ensuremath{\displaystyle #2}}}}

\makeatletter
\renewcommand*\env@matrix[1][\arraystretch]{%
  \edef\arraystretch{#1}%
  \hskip -\arraycolsep
  \let\@ifnextchar\new@ifnextchar
  \array{*\c@MaxMatrixCols c}}
\makeatother

\renewcommand{\arraystretch}{1.5}
\DeclareMathOperator*{\argmax}{arg\, max}

\author{Jian-Jia~Weng, Fady~Alajaji, and Tam{\'a}s~Linder%
\thanks{The authors are with the Department of Mathematics and Statistics, Queen’s University, Kingston, ON K7L 3N6, Canada (email: jian-jia.weng@queensu.ca, fa@queensu.ca, linder@mast.queensu.ca).}
\thanks{This work was supported in part by NSERC of Canada.}}

\newcommand{\mli}[1]{\mathit{#1}}
\begin{document}

\title{A Simple Capacity Outer Bound for Two-Way Channels and Capacity Approximation Results}

\maketitle
\begin{abstract}
  Channel symmetry properties that imply the tightness of Shannon's random coding inner bound have recently been used to determine the capacity region of discrete-memoryless two-way channels (DM-TWCs). 
  For channels without such symmetry properties, outer bounds are often needed to estimate the capacity region.
  However, validating symmetry conditions and/or evaluating non-trivial outer bounds are computationally demanding, especially for channels with large input and output alphabets. 
  In this paper, three easy-to-check conditions that identify DM-TWCs with no such symmetry properties as well as an easy-to-compute outer bound are derived.
  The bound is obtained from Shannon's inner bound computation but is non-trivial.
  Using this outer bound, approximate capacity results can be established for certain DM-TWCs. 
  The results are illustrated by two examples. 
\end{abstract}


\vspace{-0.3cm}\section{Introduction}\label{sec:introduction}
Shannon's two-way communication \cite{shannon1961} allows two terminals to operate in a full-duplex manner, which is ideally the most efficient way to utilize limited channel resources. 
To date, it is still unclear in general how a terminal can maximize its own information transmission rate and concurrently provide feedback to optimally help the other terminal's transmission. 
Because of this, the problem of finding a single-letter expression for the capacity region for general two-way channels (TWCs) remains unsolved. 
The best known capacity inner and outer bounds for the discrete-memoryless (DM) setup are given in \cite{han1984} and \cite{zhang1986}, respectively.  

Recently, various efforts have been made to investigate DM-TWCs \cite{sabag2018, gu2019,palacio2019,jjw2019isit,jjw2019cwit,jjw2019}.\footnote{A summary of earlier work on DM-TWCs can be found in \cite{meeuwissen1998}.
One can also find results on continuous TWCs \cite{han1984,varshney2013,seo2019} or multi-terminal multi-way variants \cite{ong2012, cheng2014, chaaban2016}, but these works are outside the scope of this paper.} 
A symbol-wise adaptive coding scheme was devised in \cite{sabag2018} to obtain a capacity inner bound for common-output DM-TWCs. 
Improved inner and outer bounds were also derived for the non-adaptive zero-error capacity region of general DM-TWCs \cite{gu2019}. 
These bounds subsumed the best existing results when specialized to binary-multiplying TWCs \cite[Section 13]{shannon1961}.  
Moreover, Shannon's channel symmetry property \cite[Section 11]{shannon1961}, under which his random coding inner bound is tight, was extended in \cite{chaaban2017,jjw2019cwit,jjw2019} in order to determine the exact capacity region for a larger class of DM-TWCs. 
These results suggest a procedure, shown in Section~\ref{sec:procedure}, to assess the capacity region of a general DM-TWC. 
Nevertheless, such a procedure is computationally demanding. 
It is thus of interest whether one can reduce the computational complexity associated with such procedure. 

In this paper, we derive theoretical results that help avoid unnecessary calculations during this procedure. 
Our main contribution is the derivation of a simple but non-trivial capacity outer bound. 
The bound is constructed from Shannon's inner bound calculations. 
We note that our goal is not to refine capacity outer bound results \cite{shannon1961, zhang1986}, but to seek a low-complexity, non-trivial outer bound. 
The idea is that when the low-complexity outer bound is close to Shannon's inner bound, then we can determine the capacity region with large accuracy and hence the calculation of other sophisticated outer bounds is no longer needed.  
We give examples to illustrate the capacity approximation results. 
Moreover, we derive three easy-to-check conditions to identify DM-TWCs that do not possess the symmetry properties defined in \cite[Theorem 1]{jjw2019} and \cite[Theorem 4]{jjw2019}\footnote{These two theorems currently give the most general conditions for two-terminal DM-TWCs under which Shannon's capacity inner bound is tight.} (see Propositions~\ref{prop:thm1} and~\ref{prop:thm4} in Section II).
These conditions allow us to skip the complex validation process in determining the symmetry properties in many situations. 

The rest of this paper is organized as follows. 
In Section~II, the system model and a procedure for finding the capacity region of a general DM-TWC are given.  
The computational aspects of the procedure are discussed in Section~III, together with three necessary conditions for a DM-TWC to possess desirable symmetry properties.
In Section~IV, a simple outer bound as well as capacity approximation results are presented. 
Conclusions are drawn in Section~V. 

\section{Preliminaries}\label{sec:p2p}
\begin{figure}[!tb]
    \begin{centering}
    \includegraphics[scale=0.45, draft=false]{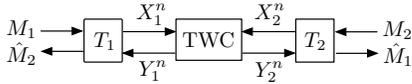}
    \caption{Block diagram of point-to-point two-way transmission.\label{fig:twcmodel}}
    \vspace{-0.5cm}
    \end{centering}
  \end{figure}

In a two-way transmission system, shown in Fig.~\ref{fig:twcmodel}, two terminals exchange messages $M_1$ and $M_2$ via $n$ uses of a shared channel. 
The messages $M_1$ and $M_2$ are assumed to be independent and uniformly distributed on the finite sets $\mathcal{M}_1$ and $\mathcal{M}_2$, respectively. 
For $j=1, 2$, let $\mathcal{X}_j$ and $\mathcal{Y}_j$ denote the finite channel input and output alphabets of terminal~$j$. 
The joint probability distribution of all random variables for $n$ uses of the channel is given by
\begin{IEEEeqnarray}{l}
\scalemath{0.84}{P_{M_1, M_2, X_1^n, X_2^n, Y_1^n, Y_2^n}=\frac{1}{|\mathcal{M}_1||\mathcal{M}_2|}\cdot\left(\prod\limits_{i=1}^n P_{X_{1,i}|M_1, Y_1^{i-1}}\right)}\nonumber\\
\qquad\ \ \scalemath{0.84}{\cdot\left(\prod\limits_{i=1}^n P_{X_{2, i}|M_2, Y_2^{i-1}}\right)\cdot\left(\prod\limits_{i=1}^n P_{Y_{1,i}, Y_{2, i}|X_1^i, X_2^i, Y_1^{i-1}, Y_2^{i-1}}\right),}\nonumber
\end{IEEEeqnarray}
where $X_j^i\triangleq (X_{j, 1}, X_{j, 2}, \dots, X_{j, i})$ and $Y_j^i\triangleq (Y_{j, 1}, Y_{j, 2}, \allowbreak\dots, \allowbreak Y_{j, i})$ for $j=1, 2$ and $i=1, 2, \dots, n$. 
The $n$ transmissions can be then described by the sequence of conditional probabilities $\{ P_{Y_{1,i},Y_{2,i}|X_{1}^i,X_{2}^i,Y_{1}^{i-1},Y_{2}^{i-1} }\}_{i=1}^n$. 
When the TWC is memoryless with transition probability $P_{Y_1, Y_2|X_1, X_2}$, we further have that $P_{Y_{1,i},Y_{2,i}|X_{1}^i,X_{2}^i,Y_{1}^{i-1},Y_{2}^{i-1}}=P_{Y_{1,i},Y_{2,i}|X_{1, i}, X_{2, i}}=P_{Y_1, Y_2|X_1, X_2}$ for all $i=1, 2, \dots, n$.
We will only consider memoryless TWCs in this paper. 

\begin{definition}
An $(n, R_1, R_2)$ channel code for the DM-TWC consists of two message sets $\mathcal{M}_1=\{1, 2, \dots, 2^{\mli{nR}_1}\}$ and $\mathcal{M}_2=\{1, 2, \dots, 2^{\mli{nR}_2}\}$, two sequences of encoding functions $\bm{f}_1\triangleq (f_{1,1}, f_{1,2}, \dots, f_{1, n})$ and $\bm{f}_2\triangleq (f_{2,1}, f_{2,2}, \dots, f_{2,n})$, where $f_{j,1}:  \mathcal{M}_j \to \mathcal{X}_j$ and $f_{j, i}:  \mathcal{M}_j \times  \mathcal{Y}_j^{i-1} \to \mathcal{X}_j$ 
for $j=1, 2$ and $i=2, 3, \dots, n$, and two decoding functions $g_1: \mathcal{M}_1 \times  \mathcal{Y}_1^{n} \to \mathcal{M}_2$ and $g_2: \mathcal{M}_2 \times  \mathcal{Y}_2^{n} \to \mathcal{M}_1$. 
\end{definition}

The channel inputs at the first time slot are only functions of the messages, i.e., $X_{j, 1} = f_{j, 1}(M_j)$ for $j=1, 2$, but all subsequent channel inputs are generated by also adapting to the previous channel outputs, i.e., $X_{j, i} = f_{j, i}(M_j, Y_j^{i-1})$ for $i=2, 3, \dots, n$.
After all $n$ channel outputs are observed, terminal~$j$ reconstructs $M_{j'}$ as $\hat{M}_{j'} =g_j(M_j, Y_j^n)$ for $j, j'=1, 2$ with $j\neq j'$. 
We define the average probability of decoding error as $P^{(n)}_{e}(\bm{f}_1, \bm{f}_2, g_1, g_2)=\text{Pr}\{\hat{M}_1 \neq M_1\ \text{or}\ \hat{M}_2 \neq M_2\}$, which leads to the following definition. 

\begin{definition}
  A rate pair $(R_1,R_2)$ is said to be achievable if there exists a sequence of $(n, R_1, R_2)$ codes such that $\lim_{n \to \infty} P^{(n)}_{e}=0$.
  The capacity region $\mathcal{C}$ of a DM-TWC is the closure of the convex hull of all achievable rate pairs. 
  \end{definition}

\vspace{-0.5cm}\subsection{Capacity Inner and Outer Bounds}
Let $\mathcal{R}(P_{X_1,X_2},P_{Y_1,Y_2|X_1,X_2})\triangleq\{(R_1,R_2): 0\le R_1\le I(X_1;Y_2|X_2), 0\le R_2 \le I(X_2;Y_1|X_1)\}$, where the joint distribution of all random variables is given by $P_{X_1, X_2, Y_1, Y_2}=P_{X_1,X_2} P_{Y_1,Y_2|X_1,X_2}$.
Let $\mathcal{P}(\mathcal{A})$ denote the set of all probability distributions on a finite set $\mathcal{A}$. 
Although a single-letter characterization of $\mathcal{C} $ remains unsolved, Shannon proved that $\mathcal{C}$ can be inner bounded by \cite{shannon1961} 
 \begin{align*}
 \scalemath{0.9}{\mathcal{C}_{\text{I}} (P_{Y_1,Y_2|X_1,X_2}) \triangleq \scalemath{0.81}{\overline{\text{co}} \left( \bigcup_{P_{X_1}\in\mathcal{P}(\mathcal{X}_1), P_{X_2}\in\mathcal{P}(\mathcal{X}_2)}\mathcal{R}\left(P_{X_1}{\cdot} P_{X_2}, P_{Y_1,Y_2|X_1,X_2}\right)\right)}},
 \end{align*}
 and outer bounded by
 \begin{align*}
\scalemath{0.9}{\mathcal{C}_{\text{O}} (P_{Y_1,Y_2|X_1,X_2}) \triangleq \scalemath{0.89}{\bigcup_{P_{X_1,X_2}\in\mathcal{P}(\mathcal{X}_1\times\mathcal{X}_2)} \mathcal{R}\left(P_{X_1,X_2}, P_{Y_1,Y_2|X_1,X_2}\right)}},
  \end{align*}
where $\overline{\text{co}}(\cdot)$ denotes taking the closure of the convex hull.

We remark that for DM-TWCs, there is a trivial outer bound given by $\tilde{\mathcal{C}}_{\text{O}}\triangleq\{(R_1, R_2):0\le R_1\le I^*_1, 0\le R_2\le I^*_2\}$, where $I^{*}_j\triangleq\max_{x_{j'}\in\mathcal{X}_{j'}}\allowbreak\max_{P_{X_j}\in\mathcal{P}({\mathcal{X}_j})}\allowbreak I(X_j; Y_{j'}|X_{j'}=x_{j'})$, $j\neq j'$, but it is loose except for some very special cases. 

\vspace{-0.2cm}\subsection{Channel Symmetry Properties}
In general, $\mathcal{C}_{\text{I}}$ and $\mathcal{C}_{\text{O}}$ do not coincide.
We here review two recently established channel symmetry properties that imply $\mathcal{C}_{\text{I}}=\mathcal{C}_{\text{O}}$ \cite{jjw2019}. 
For simplicity, we occasionally use $\mathcal{I}(P_{X}, P_{Y|X})$ to denote the mutual information $I(X; Y)$ between input $X$ (governed by $P_X$) and the corresponding output $Y$ of a channel with transition probability $P_{Y|X}$.
For $j, j'=1, 2$ with $j\neq j'$, we use superscript in $I^{(l)}(X_{j'}; Y_j|X_j)$, $H^{(l)}(Y_j|X_j)$, and $H^{(l)}(Y_j|X_1, X_2)$ to indicate that these quantities are evaluated under the joint input distribution $P^{(l)}_{X_1, X_2}=P^{(l)}_{X_j} P^{(l)}_{X_{j'}|X_j}$.  

\begin{proposition}[\hspace{1sp}{\cite[Theorem~1]{jjw2019}}]
\label{prop:thm1}
If a DM-TWC satisfies conditions (a) and (b1) below, then $\mathcal{C}_\emph{I}=\mathcal{C}_\emph{O}$. 
\begin{itemize}
\item[(a)] \underline{Common optimal input distribution}: there exists $P^*_{X_1}\in\mathcal{P}(\mathcal{X}_1)$ such that  for all $x_2\in\mathcal{X}_2$, we have that $\argmax_{P_{X_1|X_2=x_2}} I(X_1; Y_2|X_2=x_2)=P^*_{X_1};$ 
\item[(b1)] \underline{Invariance of input-output mutual information}: for any fixed $P_{X_2}\in\mathcal{P}(\mathcal{X}_2)$, we have that $\mathcal{I}(P_{X_2}, P_{Y_1|X_1=x_1, X_2})$ does not depend on $x_1\in\mathcal{X}_1$. 
\end{itemize}
\end{proposition}

\begin{proposition}[\hspace{1sp}{\cite[Theorem~4]{jjw2019}}]
\label{prop:thm4}
If a DM-TWC satisfies condition (a) above and condition (b2) below, then $\mathcal{C}_\emph{I}=\mathcal{C}_\emph{O}$:
\begin{itemize}
\item[(b2)] $H(Y_1|X_1, X_2)$ does not depend on $P_{X_1|X_2}$ given $P_{X_2}$, and $H^{(1)}(Y_1|X_1)\le H^{(2)}(Y_1|X_1)$ holds for any $P^{(1)}_{X_1, X_2}=P^{(1)}_{X_2} P^{(1)}_{X_1|X_2}$ and $P^{(2)}_{X_1, X_2}=P^*_{X_1} P^{(1)}_{X_2}$.
\end{itemize}
\end{proposition}

Given the conditions of either propositions, one can deduce that $I^{(1)}(X_{j'}; Y_j|X_j)\le I^{(2)}(X_{j'}; Y_j|X_j)$, $j\neq j'$, for any $P^{(1)}_{X_1, X_2}=P^{(1)}_{X_2}P^{(1)}_{X_1|X_2}$ and $P^{(2)}_{X_1, X_2}=P^{*}_{X_1}P^{(1)}_{X_2}$, thus implying that $\mathcal{C}_{\text{O}}\subseteq\mathcal{C}_{\text{I}}$.
A crucial implication of the equality $\mathcal{C}_{\text{I}}=\mathcal{C}_{\text{O}}$ is that using independent inputs can achieve capacity and hence adaptive coding is unnecessary. 

\vspace{-0.3cm}
\subsection{A Procedure for Finding the Capacity Region}\label{sec:procedure}
Combining the known results and taking the computational complexity into accounts, we suggest the following procedure to assess the capacity region of a general DM-TWC: 
\begin{itemize}[leftmargin=+1.4cm]
  \item[Step 1:] Validate channel symmetry conditions, e.g., Propositions~\ref{prop:thm1} or~\ref{prop:thm4} or other more restrictive results in \cite{jjw2019};
  \item[Step 2:] If the channel possesses desirable symmetry properties, then we compute $\mathcal{C}_{\text{I}}$; otherwise, evaluate capacity inner and outer bounds\footnote{
    In practice, Shannon's capacity bounds $\mathcal{C}_{\text{I}}$ and $\mathcal{C}_{\text{O}}$ are commonly used to infer the capacity region since they are easier to evaluate than other capacity bounds \cite{han1984,zhang1986}.} and compare them.  
\end{itemize}

To further reduce the computational demand of each step, in Section~III we seek necessary conditions for a DM-TWC to satisfy the symmetry properties in Propositions~1 and~2. We also derive a simple but non-trivial capacity outer bound.


\vspace{-0.2cm}
\section{Computation-Reduction Strategy for Validating Symmetry Conditions}
This section first highlights which part of the procedure may involve intensive computations. 
Then, we derive three easy-to-check conditions to avoid the complex validation step.  

%
\vspace{-0.4cm}\subsection{Computational Complexity}\label{sec:compissue}
Known numerical methods to obtain Shannon's capacity bounds involve (i) uniformly quantizing the probability simplex of channel inputs; (ii) evaluating the region $\mathcal{R}$ for each quantized input distribution; (iii) taking the convex hull. 
By definition, the quantized input distributions for $\mathcal{C}_{\text{I}}$ and $\mathcal{C}_{\text{O}}$ take values in the spaces $\mathcal{P}(\mathcal{X}_1)\times \mathcal{P}(\mathcal{X}_2)$ and $\mathcal{P}(\mathcal{X}_1\times\mathcal{X}_2)$, with dimensions $(|\mathcal{X}_1|-1)(|\mathcal{X}_2|-1)$ and $|\mathcal{X}_1||\mathcal{X}_2|-1$, respectively. 
Let $\Delta\in (0, 1)$ denote the step size of the quantization and suppose that $\Delta^{-1}\in\mathbb{N}$. 
We thus have ${\Delta^{-1}+|\mathcal{X}_1|-1 \choose \Delta^{-1}}{\Delta^{-1}+|\mathcal{X}_2|-1 \choose \Delta^{-1}}$ and ${\Delta^{-1}+|\mathcal{X}_1||\mathcal{X}_2|-1 \choose \Delta^{-1}}$ quantized input distributions to compute for $\mathcal{C}_{\text{I}}$ and $\mathcal{C}_{\text{O}}$, respectively.  
For example, when all channel alphabet sizes are not larger than $3$, setting $\Delta=0.025$ is enough to have visually indistinguishable region estimates of $\mathcal{C}_{\text{I}}$ and $\mathcal{C}_{\text{O}}$. 
The number of quantized input distributions for $\mathcal{C}_{\text{I}}$ is roughly $7\times 10^{6}$, but it is about $3\times 10^9$ for $\mathcal{C}_{\text{O}}$; evaluating $\mathcal{C}_{\text{O}}$ clearly involves significantly more calculations.
Although one can apply the Lagrange multiplier method \cite[Section 11]{shannon1961} to find $\mathcal{C}_{\text{O}}$, the implementation cost is still considerable. 


On the other hand, even though the validation of individual symmetry conditions is usually easy, the accumulated computational complexity can be significant.  
For instance, validating condition~(a) can be efficiently done via the Blahut-Arimoto algorithm \cite{blahut1972}. 
The verification of condition (b1), though slightly complex, only involves checking input distributions in $\mathcal{P}({\mathcal{X}_2})$. 
Clearly, the overall computational complexity for Proposition~\ref{prop:thm1} is lower than the one needed for evaluating $\mathcal{C}_{\text{O}}$. 
Nevertheless, if this validation fails, then we need to swap the roles of terminals~1 and~2 and verify conditions (a) and (b1) again. 
If this process is still unsuccessful, then one may switch to Proposition~\ref{prop:thm4}. 
Hence, the entire validation process can be lengthy and requires significant computational resources. 

\vspace{-0.2cm}\subsection{Necessary Conditions for Symmetric Channels}
To reduce the computational complexity of validating symmetry conditions, we provide three simple conditions that can be used to rule out such symmetry properties. 
Two of these conditions appeared in the proofs of \cite[Corollary 2]{jjw2019} and \cite[Theorem 7]{jjw2019}.
As the conditions are useful in practice, we present them here as standalone results without proof. 
The first one is derived for condition (b1) of Proposition~\ref{prop:thm1}, which can be efficiently validated via the Blahut-Arimoto algorithm (as done for condition (a)). 
The second one is for condition (b2) of Proposition~\ref{prop:thm4}, which sometimes can be verified by directly observing the channels' marginal transition matrices. 
\begin{theorem}
  \label{thm:nc1}
  If a DM-TWC satisfies condition (b1) of Proposition~\ref{prop:thm1}, then there exists $P^*_{X_2}\in\mathcal{P}(\mathcal{X}_2)$ such that for all $x_1\in\mathcal{X}_1$, $\argmax_{P_{X_2|X_1=x_1}} I(X_2; Y_1|X_1=x_1)=P^*_{X_2}$. 
\end{theorem}


\begin{theorem}
  \label{thm:nc2}
  If a DM-TWC satisfies condition (b2) of Proposition~\ref{prop:thm4}, then $H(Y_1|X_1=x'_1, X_2=x_2)=H(Y_1|X_1=x''_1, X_2=x_2)$ for any $x'_1, x''_1\in\mathcal{X}_1$ and fixed $x_2\in\mathcal{X}_2$.
\end{theorem}

Furthermore, for DM-TWCs that satisfy the conditions of Proposition~\ref{prop:thm4}, the inequality $I^{(1)}(X_{j'}; Y_j|X_j)\le I^{(2)}(X_{j'}; Y_j|X_j)$, $j\neq j'$, holds for the specific input distributions $P^{(1)}_{X_1, X_2}\allowbreak = P^{*}_{X_1} P^{(1)}_{X_2|X_1}=P^{(1)}_{X_2}P^{(1)}_{X_1|X_2}$ and $P^{(2)}_{X_1, X_2}=P^{*}_{X_1} P^{(1)}_{X_2}$. 
For $j{=}2$ and $j’{=}1$, expanding the inequality, we have that $\sum_{x_1\in\mathcal{X}_1}P^{*}_{X_1}(x_1)\cdot\mathcal{I}\big(P^{(1)}_{X_2|X_1=x_1}, P_{Y_1|X_1=x_1, X_2}\big)\le \sum_{x_1\in\mathcal{X}_1}P^{*}_{X_1}(x_1)\cdot\mathcal{I}\big(P^{(1)}_{X_2}, P_{Y_1|X_1=x_1, X_2}\big)$,
which indicates that using an \emph{average input} $P^{(1)}_{X_2}$ at terminal~$2$ does not incur any information loss when terminal~$1$ uses the common optimal input $P^{*}_{X_1}$. 
As one can choose $P^{(1)}_{X_2|X_1=x_1}=\argmax_{P_{X_2|X_1=x_1}}\allowbreak \mathcal{I}(P_{X_2|X_1=x_1}, P_{Y_1|X_1=x_1, X_2})$, the inequality suggests another common maximizer property, which we use to derive a necessary condition for (a) and (b2).

\begin{theorem}
  \label{thm:necondgeneralizedCVA}
  If a DM-TWC satisfies the conditions of Proposition~\ref{prop:thm4} with $P^{*}_{X}(x_1)>0$ for all $x_1\in\mathcal{X}_1$, then there exists a common output conditional entropy maximizer $P^*_{X_2}$ such that for all $x_1\in\mathcal{X}_1$, 
  $\argmax_{P_{X_2|X_1=x_1}} H(Y_1|X_1=x_1) = P^{*}_{X_2}$.  
\end{theorem}
\vspace{-0.1cm}\begin{IEEEproof}
  For any $P^{(1)}_{X_1, X_2}=\allowbreak P^{(1)}_{X_2} P^{(1)}_{X_1|X_2}=P^{(1)}_{X_1} P^{(1)}_{X_2|X_1}$, let $P^{(2)}_{X_1, X_2}=P^*_{X_1} P^{(1)}_{X_2}$; the symmetry condition (b2) gives the inequality $H^{(1)}(Y_1|X_1)\le H^{(2)}(Y_1|X_1)$. 
  Consider the particular choice $P^{(1)}_{X_1}=P^{*}_{X_1}$ and $P^{(1)}_{X_2|X_1=x_1}=\allowbreak\argmax_{P_{X_2|X_1=x_1}} \allowbreak H(Y_1|X_1=x_1)$ for all $x_1\in\mathcal{X}_1$.
  Together with the assumption that $P^{*}_{X_1}(x_1)>0$ for all $x_1\in\mathcal{X}_1$ and the non-negativity of entropy, we obtain that $H^{(1)}(Y_1|X_1=x_1)\le \allowbreak H^{(2)}(Y_1|X_1=x_1)$ for every $x_1{\in}\mathcal{X}_1$.
  Since $P^{(1)}_{X_2|X_1=x_1}$ is the maximizer of $H(Y_1|X_1=x_1)$, we further have that $H^{(1)}(Y_1|X_1=x_1)=\allowbreak H^{(2)}(Y_1|X_1=x_1)$ for any $x_1\in\mathcal{X}_1$.
  In other words, the conditional entropies $H(Y_1|X_1=x_1)$, $x_1\in\mathcal{X}_1$, have a common maximizer (and $P^{(1)}_{X_2}$ is the common maximizer). 
\end{IEEEproof}

As $H(Y_1|X_1=x_1)$ is a concave function of $P_{X_2|X_1=x_1}$ for fixed $P_{Y_1|X_1=x_1, X_2}$, one can use a standard convex optimization program to check this necessary condition.  
This result is useful since the inequality $H^{(1)}(Y_1|X_1)\le \allowbreak H^{(2)}(Y_1|X_1)$ in condition (b2) is often difficult to verify.

Due to space limitation, we present herein a single example. 
Given a binary-input and binary-output DM-TWC, where the conditional distributions $[P_{Y_1|X_1, X_2}(\cdot|x_1, \cdot)]$ and $[P_{Y_2|X_1, X_2}(\cdot|\cdot, x_2)]$ correspond to binary symmetric channels (BSCs) but with different crossover probabilities,   
condition (a) of Proposition~\ref{prop:thm4} is satisfied since the uniform input distribution is optimal for all BSCs. However, condition (b2) fails to hold due to Theorem~\ref{thm:nc2} and hence the validation of condition (b2) is no longer needed.  
Furthermore, although Theorem~\ref{thm:nc1} holds in this example, condition (b1) does not hold (since a fixed input distribution cannot yield the same input-output mutual information for BSCs with distinct crossover probabilities). 



\vspace{-0.2cm}\section{A Simple Outer Bound to the Capacity Region}
For DM-TWCs that are not symmetric in the sense of Propositions~\ref{prop:thm1} or~\ref{prop:thm4}, calculating outer bounds to assess the capacity region is almost inevitable. 
The discussion in Section~\ref{sec:compissue} highlighted the high computational demands for obtaining $\mathcal{C}_{\text{O}}$. 
When using refined outer bounds such as in \cite{zhang1986}, the problem becomes even more complex due to the use of auxiliary random variables. 
In this section, we derive an easy-to-compute but non-trivial outer bound.
Also, we give examples where our bound (together with $\mathcal{C}_{\text{I}}$) results in a good estimate of $\mathcal{C}$. 
We note that our goal here is not to improve on any outer bound results; instead, we show that the computation of $\mathcal{C}_{\text{I}}$ can in itself produce a useful outer bound.

Our result is inspired by the proof of Proposition~\ref{prop:thm1}. 
To derive our simple outer bound, we relax the symmetry conditions of Proposition~\ref{prop:thm1} as follows. 
Without loss of generality, the definitions are given for a specific direction of transmission.

\begin{definition}
  \label{def:amax}
Given the set of state-dependent one-way channels $\{P_{Y_2|X_1, X_2}(\cdot|\cdot, x_2):x_2\in\mathcal{X}_2\}$, let 
\vspace{-0.1cm}\begin{IEEEeqnarray}{l}
\scalemath{0.85}{\alpha^*=\min_{\tilde{P}_{X_1}\in\mathcal{P}(\mathcal{X}_1)}\max_{x_2\in\mathcal{X}_2}\Big|\mathcal{I}(\tilde{P}_{X_1}, P_{Y_2|X_1, X_2=x_2})}
\nonumber\\ [-0.27cm]
\qquad\qquad\qquad\qquad\qquad \scalemath{0.9}{-\max_{P_{X_1}\in\mathcal{P}(\mathcal{X}_1)}\mathcal{I}(P_{X_1}, P_{Y_2|X_1, X_2=x_2})\Big|.}\IEEEeqnarraynumspace\*\label{eq:epsmaximizer}
\end{IEEEeqnarray}
Such a collection of channels is said to have an $\alpha^*$-close common optimal input distribution. 
\end{definition} 

\begin{remark}
  When $|\mathcal{X}_1|=2$, we have that $\alpha^*\le 0.011$ \cite{shulman2004} for \emph{any} finite collection of memoryless one-way channels under the uniform input: $\tilde{P}_{X_1}(0)=\tilde{P}_{X_1}(1)=1/2$.
\end{remark}

\begin{definition}
  \label{def:binv}
Given the set of state-dependent one-way channels $\{P_{Y_1|X_1, X_2}(\cdot|x_1, \cdot): x_1\in\mathcal{X}_1\}$, let 
  \vspace{-0.1cm}\begin{IEEEeqnarray}{l}
    \scalemath{0.85}{\beta^*={\max_{P_{X_2}\in\mathcal{P}(\mathcal{X}_2)}}\max_{x_1, x'_1\in\mathcal{X}_1\atop x_1\neq x'_1}\Big|\mathcal{I}(P_{X_2}, P_{Y_1|X_1=x_1, X_2})}\nonumber\\[-0.27cm]
    \qquad\qquad\qquad\qquad\qquad\qquad\qquad  \scalemath{0.9}{- \mathcal{I}(P_{X_2}, P_{Y_2|X_1=x'_1, X_2})\Big|.} \IEEEeqnarraynumspace
  \label{eq:epsinvariantmi}
  \end{IEEEeqnarray}
Such a collection of channels is said to be $\beta^*$-invariant in the input-output mutual information.  
\end{definition}
 
Based on Definitions~\ref{def:amax}-\ref{def:binv}, we obtain the following lemma, which will be used to obtain our capacity outer bound result. 
\begin{lemma}
  \label{thm:nearsym}
  For any DM-TWC and any achievable rate pair $(R_1, R_2)$, there exists a rate pair  $(R'_1, R'_2)$ in $\mathcal{C}_{\textnormal{I}}$ such that $R_1\le R'_1+\alpha^*$ and $R_2\le R'_2+2\beta^*$.  
\end{lemma}
  \vspace{-0.1cm}\begin{IEEEproof}
  Given any $P^{(1)}_{X_1, X_2}=P^{(1)}_{X_2} P^{(1)}_{X_1|X_2}$, let $P^{(2)}_{X_1, X_2}\triangleq P^{\alpha^*}_{X_1} P^{(1)}_{X_2}$, where $P^{\alpha^{*}}_{X_1}$ denotes the $\alpha^*$-close common optimal input distribution. First, we have that
  \begin{IEEEeqnarray}{l}
  I^{(1)}(X_1; Y_2|X_2)\nonumber\\
  \ \ = \sum_{x_2\in\mathcal{X}_2}P^{(1)}_{X_2}(x_2)\cdot \mathcal{I}\big(P^{(1)}_{X_1|X_2=x_2}, P_{Y_2|X_1, X_2=x_2}\big)\nonumber\\
  \ \ \le  \sum_{x_2\in\mathcal{X}_2}P^{(1)}_{X_2}(x_2)\cdot \bigg(\mathcal{I}\big(P^{\alpha^{*}}_{X_1}, P_{Y_2|X_1, X_2=x_2}\big)+\alpha^{*}\bigg)\nonumber\\
  \ \ = \sum_{x_2\in\mathcal{X}_2}P^{(1)}_{X_2}(x_2)\cdot \mathcal{I}\big(P^{\alpha^{*}}_{X_1}, P_{Y_2|X_1, X_2=x_2}\big) + \alpha^{*}\nonumber\\
  \ \ = I^{(2)}(X_1; Y_2|X_2) + \alpha^{*},\nonumber
  \end{IEEEeqnarray}
  where the inequality follows from Definition~\ref{def:amax}.
  Moreover, 
  \begin{IEEEeqnarray}{l}
  I^{(1)}(X_2; Y_1|X_1)\nonumber\\ 
  \ \ = \sum_{x_1\in\mathcal{X}_1}P^{(1)}_{X_1}(x_1)\cdot \mathcal{I}\big(P^{(1)}_{X_2|X_1=x_1}, P_{Y_1|X_1=x_1, X_2}\big)\nonumber\\
  \ \ \le \sum_{x_1\in\mathcal{X}_1}P^{(1)}_{X_1}(x_1)\cdot \scalemath{0.9}{\left(\mathcal{I}\left(P^{(1)}_{X_2|X_1=x_1}, P_{Y_1|X_1=x'_1, X_2}\right)+\beta^{*}\right)}\nonumber\\
  \ \ \le  \mathcal{I}\big(P^{(1)}_{X_2}, P_{Y_1|X_1=x'_1, X_2}\big)+\beta^{*}\nonumber\\ 
  \ \ =  \scalemath{0.9}{\Bigg(\sum_{x_1\in\mathcal{X}_1}P^{\alpha^{*}}_{X_1}(x_1)\Bigg)}\cdot\mathcal{I}\big(P^{(1)}_{X_2}, P_{Y_1|X_1=x'_1, X_2}\big)+\beta^{*}\nonumber\\
  \ \ \le  \sum_{x_1\in\mathcal{X}_1}P^{\alpha^{*}}_{X_1}(x_1)\cdot\scalemath{0.9}{\bigg(\mathcal{I}\big(P^{(1)}_{X_2}, P_{Y_1|X_1=x_1, X_2}\big)+\beta^{*}\bigg)}+\beta^{*}\nonumber\\
  \ \ = I^{(2)}(X_2; Y_1|X_1) + 2\beta^{*},\nonumber
  \end{IEEEeqnarray}
  where the first and the last inequalities are due to Definition~\ref{def:binv} while the second inequality holds since $\mathcal{I}(\cdot, \cdot)$ is concave in the first argument. 
  The claim is proved by noting that $(R'_1, R'_2)=(I^{(2)}(X_1; Y_2|X_2), I^{(2)}(X_2; Y_1|X_1))\in\mathcal{C}_{\text{I}}$.
\end{IEEEproof}

\begin{theorem}
  \label{thm:outerbound}
  For any DM-TWC,
  \begin{IEEEeqnarray}{l}
    \scalemath{0.9}{\mathcal{C}\subseteq\overline{\textnormal{co}}\Bigg(\bigcup_{(R'_1, R'_2)\in\mathcal{C}_{\textnormal{I}}}\{(R'_1, R'_2)\}\cup\{(R'_1+\alpha^*, R'_2+2\beta^*)\}\Bigg)\cap \tilde{C}_{\textnormal{O}}},\nonumber
  \end{IEEEeqnarray}
  where $\tilde{C}_{\textnormal{O}}$ is the trivial outer bound defined in Section II-A.
\end{theorem}

The proof of the Theorem~\ref{thm:outerbound} is omitted since it is straightforward given Lemma~\ref{thm:nearsym}. 
We emphasize that $\alpha^*$ merely depends on the marginal input distributions on $\mathcal{X}_1$ (i.e., $P_{X_1}$ and $\tilde{P}_{X_1}$) and the marginal channel distribution $P_{Y_2|X_1, X_2}$. 
More importantly, the mutual information quantities in \eqref{eq:epsmaximizer} are already found when computing $\mathcal{C}_{\text{I}}$. 
One can thus efficiently obtain $\alpha^*$ within the framework of Shannon's inner bound computation; the same holds for $\beta^*$.
As a result, forming this outer bound only requires subtraction and comparison operations.


Moreover, our outer bound coincides with $\mathcal{C}_\text{O}$ when $\alpha^*=\beta^*=0$, which is exactly equal to $\mathcal{C}_{\text{I}}$ as can be deduced from the proof of Lemma~\ref{thm:nearsym} (and hence recovers the result in Proposition~\ref{prop:thm1}). 
For other cases, the values $\alpha^*$ and $2\beta^*$ roughly indicate how much our outer bound deviates from $\mathcal{C}_{\text{I}}$ in the $R_1$ and $R_2$ axis, respectively.  
Using this fact, we can establish approximation capacity results for small deviations.  

\begin{definition}
  \label{def:epscap}
For $\epsilon\ge 0$, the $\epsilon$-neighborhood $\mathcal{C}_{\emph{I}, \epsilon}$ of Shannon's inner bound is defined as 
$\mathcal{C}_{\emph{I}, \epsilon}\triangleq\{(R_1, R_2)\in [0, I^*_1]\times [0, I^*_2]: \max\big(\frac{|R_1-R'_1|}{I_1^*}, \frac{|R_2-R'_2|}{I_2^*}\big){\le}\ \epsilon\ \text{for some}\ (R'_1, R'_2)\in\mathcal{C}_{\emph{I}}\}.$
If $\mathcal{C}_{\emph{O}}\subseteq\mathcal{C}_{\emph{I}, \epsilon}$, then $\mathcal{C}_{\emph{I}, \epsilon}$ is called an $\epsilon$-approximated capacity region.
\end{definition}

Combining this definition with Lemma~\ref{thm:nearsym}, our outer bound is an $\epsilon$-approximated capacity region with $\epsilon=\max\big(\frac{\alpha^*}{I^*_1}, \frac{2\beta^*}{I^*_2}\big)$. 
Thus, a smaller value of $\epsilon$ gives an approximation $\mathcal{C}_{\text{I}, \epsilon}\approx\mathcal{C}$ with higher accuracy.
To end this section, we illustrate $\mathcal{C}_{\text{I}, \epsilon}$ via two examples. 
Example~\ref{ex:forremark} also illustrates Remark~1; Example~\ref{ex:discuss} shows a general interplay between $\mathcal{C}_{\text{I}, \epsilon}$ and the underlying channel parameters.


\begin{table}[!t]
  \caption{Marginal channel transition matrices of Example~1}
  \centering
  \scalebox{0.8}{
  \hspace{+0.2cm}
  \begin{subtable}{.25\textwidth}
    \begin{tabular}{c||c|c|c}
      $P_{Y_1|X_1, X_2}$ & $0$ & $1$ & $2$ \\ 
      \hline \hline 
      $(0, 0)$ & $0.8$ & $0.1$ & $0.1$ \\ 
      \hline 
      $(0, 1)$ & $0.1$ & $0.8$ & $0.1$ \\ 
      \hline 
      $(0, 2)$ & $0.1$ & $0.1$ & $0.8$ \\ 
      \hline \hline 
      $(1, 0)$ & $0.8$ & $0.1$ & $0.1$ \\ 
      \hline 
      $(1, 1)$ & $0.1$ & $0.8$ & $0.1$ \\ 
      \hline 
      $(1, 2)$ & $0.1$ & $0.1$ & $0.8$ \\ 
    \end{tabular}%
  \end{subtable}%
  \ \ \hfill
  \begin{subtable}{.25\textwidth}
    \begin{tabular}{c||c|c}
      $P_{Y_2|X_1, X_2}$ & $0$ & $1$ \\ 
      \hline \hline 
      $(0, 0)$ & $0.7$ & $0.3$ \\ 
      \hline 
      $(1, 0)$ & $0.1$ & $0.9$ \\ 
      \hline \hline 
      $(0, 1)$ & $1$ & $0$\\ 
      \hline 
      $(1, 1)$ & $0.25$ & $0.75$\\ 
      \hline  \hline
      $(0, 2)$ & $0.5$ & $0.5$\\ 
      \hline 
      $(1, 2)$ & $0$ & $1$\\  
    \end{tabular}
  \end{subtable}} 
  \label{tble:ex1}
  \vspace{-0.55cm}
\end{table}

\begin{figure}[t!]\vspace{-0.3cm}  
  \centering
  \includegraphics[scale=0.315, draft=false]{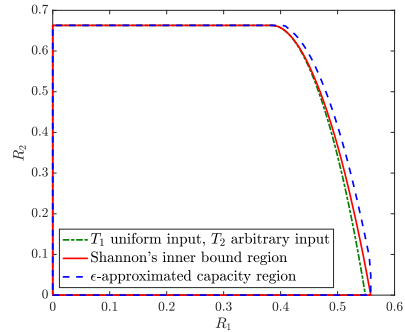}
  \caption{Numerical results for Example~\ref{ex:forremark}, where $\epsilon=0.0183$.}
  \label{fig:ex1}
  \vspace{-0.4cm}
\end{figure}

\begin{example}\vspace{-0.1cm}
  \label{ex:forremark}
Consider a DM-TWC with marginal channels given in Table~\ref{tble:ex1}. 
The channel $P_{Y_1|X_1, X_2}$ consists of two ternary sub-channels that satisfy condition (b1). 
The channel $P_{Y_2|X_1, X_2}$ is chosen to violate condition (a), which includes one Z-type, one inverse Z-type, and one pure asymmetric binary sub-channel; these sub-channels favor different input distributions. 
Based on the numerical computation for $\mathcal{C}_{\text{I}}$, we obtain that $I_1^*=0.5582$, $I_2^*=0.6603$, $\alpha^*=0.0102$, $\beta^*=0$, and an outer bound $\mathcal{C}_{\text{I}, \epsilon}$ with $\epsilon=0.0183$. 
As shown in Fig.~\ref{fig:ex1}, the region $\mathcal{C}_{\text{I}}$ and $\mathcal{C}_{\text{I}, \epsilon}$ are quite close to each other, thus providing a good estimation to $\mathcal{C}$. 
Also, for any fixed $R_2$, the rate loss of $R_1$ is less than $0.011$ (bits per channel use) when terminal~1 always uses the uniform inputs.
\end{example}

\begin{figure}[t!]\vspace{-0.12cm}
  \centering
  \includegraphics[scale=0.315, draft=false]{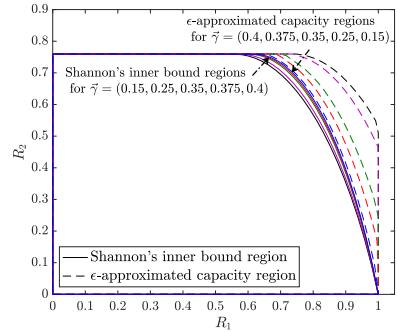}
  \caption{Numerical results for Example~\ref{ex:discuss}. The five inner regions are the $\mathcal{C}_{\text{I}}$'s corresponding to the channel parameters $\gamma=0.15$, $0.25$, $0.35$, $0.375$, and $0.4$, respectively, while the five outer regions are the corresponding $\mathcal{C}_{\text{I}, \epsilon}$'s (in reverse order).}
  \label{fig:discuss}
  \vspace{-0.4cm}
\end{figure}

\begin{example}
  \label{ex:discuss}
  Consider the DM-TWC with the marginal channels given in Table~\ref{tble:ex3}. 
  The marginal channel $P_{Y_1|X_1, X_2}$ does not satisfy condition (b1), and its sub-channels are given by perturbing a BSC with crossover probability $0.04$. 
  To demonstrate how $\mathcal{C}_{\text{I}, \epsilon}$ generally approximates $\mathcal{C}$, we also consider non-standard sub-channels for $P_{Y_2|X_1, X_2}$ parameterized by $\gamma\in [0, 0.8]$. 
  Note that when $\gamma$ increases from $0.1$ to $0.4$, the sub-channel with transition matrix $[P_{Y_2|X_1, X_2}(\cdot|\cdot, 1)]$ becomes less noisy and the overall marginal channel $P_{Y_2|X_1, X_2}$ tends to be symmetric in the sense of condition (a). 
  For this setup, we have that $\beta=0.0025$, $I_1^*=1$, and $I_2^*=0.7577$. 
  In Table~\ref{tble:discuss}, we list the values of $\alpha^*$ and $\epsilon$ for different values of $\gamma\in [0, 0.4]$. 
  We also depict $\mathcal{C}_{\text{I}}$ and $\mathcal{C}_{\text{I}, \epsilon}$ for selected values of $\gamma$ in Fig.~\ref{fig:discuss}. 
  It is observed that when $\epsilon< 0.05$, our simple outer bound and $\mathcal{C}_{\text{I}}$ determine the capacity region $\mathcal{C}$ with large accuracy. 
  In other cases, our outer bound is still non-trivial.

\end{example}

\vspace{-0.4cm}\section{Conclusions}
Two computation-reduction strategies for assessing the capacity region of general DM-TWCs were presented.  
The derived necessary conditions can quickly identify DM-TWCs that do not have the symmetry properties of \cite{jjw2019}, thus avoiding lengthy validation steps. 
Our simple but non-trivial outer bound result also eliminates the need of complex outer bound evaluations in some cases. 
Future research directions include using channel ordering to predict the capacity region and determining the exact capacity region via channel decomposition.

\begin{table}[!t]
  \caption{Marginal channel transition matrices of Example~\ref{ex:discuss}}  
  \scalebox{0.8}{
  \centering\hspace{+0.15cm}
  \begin{subtable}{.25\textwidth} 
    \begin{tabular}{c||c|c}
      $P_{Y_1|X_1, X_2}$ & $0$ & $1$ \\ 
      \hline \hline 
      $(0, 0)$ & $0.96$ & $0.04$ \\ 
      \hline 
      $(0, 1)$ & $0.04$ & $0.96$ \\ 
      \hline \hline 
      $(1, 0)$ & $0.961$ & $0.039$\\ 
      \hline 
      $(1, 1)$ & $0.041$ & $0.959$\\ 
      \hline  \hline
      $(2, 0)$ & $0.96$ & $0.04$\\ 
      \hline 
      $(2, 1)$ & $0.041$ & $0.959$\\  
    \end{tabular}%
  \end{subtable}%
  \hfill
  \begin{subtable}{.25\textwidth} 
    \begin{tabular}{c||c|c|c}
      $P_{Y_2|X_1, X_2}$ & $0$ & $1$ & $2$ \\ 
      \hline \hline 
      $(0, 0)$ & $1$ & $0$ & $0$ \\ 
      \hline 
      $(1, 0)$ & $0$ & $1$ & $0$ \\ 
      \hline 
      $(2, 0)$ & $0.5$ & $0.5$ & $0$ \\ 
      \hline \hline 
      $(0, 1)$ & $0$ & $0.1$ & $0.9$ \\ 
      \hline 
      $(1, 1)$ & $0.2$ & $\gamma$ & $0.8-\gamma$ \\ 
      \hline 
      $(2, 1)$ & $0.2$ & $0.8-\gamma$ & $\gamma$ \\ 
    \end{tabular}%
  \end{subtable}}%
  \label{tble:ex3}
  \vspace{-0.1cm}
\end{table}

\begin{table}[!t]
  \caption{The values of $\alpha^*$ and $\epsilon$ under different settings of $\gamma$ in Example~\ref{ex:discuss}}  
  \scalebox{0.8}{
  \centering
  \begin{subtable}{.5\textwidth} 
    \begin{tabular}{c||c|c|c|c|c|c|c|c}
      $\gamma$ & $0.1$ & $0.15$ & $0.2$ & $0.25$ & $0.3$ & $0.35$ & $0.375$ & $0.4$\\ 
      \hline \hline 
      $\alpha^*$ & $0.1911$ & $0.1808$ & $0.1641$ & $0.1398$ & $0.1063$ & $0.0608$ & $0.0013$& $0.001$ \\ 
      \hline 
      $\epsilon$ & $0.1911$ & $0.1808$ & $0.1641$ & $0.1398$ & $0.1063$ & $0.0608$ & $0.0325$ & $0.0066$\\ 
    \end{tabular}%
  \end{subtable}}%
  \label{tble:discuss}
  \vspace{-0.3cm}
\end{table}

\end{document}